\documentclass{article}
\usepackage{spconf,amsmath,graphicx}
\usepackage{hyperref}
\usepackage{booktabs}
\usepackage[font={small,stretch=1}]{caption}

\usepackage{enumitem}
\usepackage{float}
\setlist{nosep, leftmargin=14pt}

\usepackage{mwe} 

\title{A lightweight, rapid and efficient deep convolutional network for chest X-ray tuberculosis detection}
\name{\parbox{\linewidth}{\centering
Daniel Capellán-Martín\textsuperscript{1,2},
Juan J. Gómez-Valverde\textsuperscript{1,2},
David Bermejo-Peláez\textsuperscript{1,2},
María J. Ledesma-Carbayo\textsuperscript{1,2}}}

\address{\parbox{\linewidth}{\centering\fontsize{9.5}{10}\selectfont
\textsuperscript{1} Biomedical Image Technologies, ETSI Telecomunicación, Universidad Politécnica de Madrid \newline
\textsuperscript{2} CIBER-BBN, Instituto Salud Carlos III, Madrid, Spain
}}

\begin{document}
%

\maketitle
\begin{abstract}
\noindent
Tuberculosis (TB) is still recognized as one of the leading causes of death worldwide. Recent advances in deep learning (DL) have shown to enhance radiologists' ability to interpret chest X-ray (CXR) images accurately and with fewer errors, leading to a better diagnosis of this disease. However, little work has been done to develop models capable of diagnosing TB that offer good performance while being efficient, fast and computationally inexpensive. In this work, we propose LightTBNet, a novel lightweight, fast and efficient deep convolutional network specially customized to detect TB from CXR images. Using a total of 800 frontal CXR images from two publicly available datasets, our solution yielded an accuracy, F1 and area under the ROC curve (AUC) of 0.906, 0.907 and 0.961, respectively, on an independent test subset. The proposed model demonstrates outstanding performance while delivering a rapid prediction, with minimal computational and memory requirements, making it highly suitable for deployment in handheld devices that can be used in low-resource areas with high TB prevalence. Code publicly available at: \href{https://github.com/dani-capellan/LightTBNet}{https://github.com/dani-capellan/LightTBNet}.
\end{abstract}
\section{Introduction}
\label{sec:intro}

Tuberculosis (TB) remains one of the leading causes of death worldwide \cite{WorldHealthOrganization2022GlobalReport}. About a quarter of the world's population is infected with \textit{Mycobacterium tuberculosis}, the agent that causes TB, usually affecting the lungs, although it can also affect other parts of the body \cite{Ravimohan2018TuberculosisPathophysiology}. Prior to the coronavirus (COVID-19) outbreak, TB was the leading infectious disease-related cause of death due to a single infectious agent, surpassing HIV/AIDS \cite{WorldHealthOrganization2022GlobalReport, Koch2018MycobacteriumTuberculosis}. Since then, there has been a significant drop in the number of people newly diagnosed and treated that has led to an increase in TB deaths \cite{WorldHealthOrganization2022GlobalReport}. Early diagnosis of TB is essential to promote effective treatment and to reduce further transmission. Chest X-rays (CXRs) are recommended by the World Health Organization (WHO) as a TB screening and triage tool given their wide availability, rapid execution, and acquisition with portable machines to reduce the risk of cross infection \cite{WorldHealthOrganization2016ChestApproaches, Orsi2020FeasibilityCOVID-19}. Artificial intelligence (AI)-powered computer-aided detection (CAD) tools have shown to improve health outcomes, especially in under-resourced areas with a high TB burden \cite{Kulkarni2020ArtificialReview, Qin2021AImplementers}. By using these solutions, radiologists are able to interpret images with greater accuracy and with fewer errors, allowing them to devote more time to patient care \cite{McBee2018DeepRadiology}.

Several studies have explored the use of deep learning (DL) for TB detection in CXRs. Hwang et al.\@ \cite{Hwang2016ANetworks} and Islam et al.\@ \cite{Islam2017AbnormalityNetworks} proposed DL-based methods combining Convolutional Neural Networks (CNNs) and Transfer Learning techniques, and also explored the ensembling of predictions from multiple architectures to improve TB detection. However, efficiency is crucial when designing DL models for medical tasks, given the limited amounts of data that may lead to overfitting and poor generalization. Pasa et al.\@ \cite{Pasa2019EfficientVisualization} proposed a novel Residual Neural Network (ResNet) for TB detection, emphasizing the importance of efficiency and the need for models that can generalize well with limited data. Rajpurkar et al.\@\cite{Rajpurkar2020CheXaid:HIV} combined clinical information with CXR images in a deep learning algorithm based on a DenseNet-121 architecture to diagnose TB. Wong et al.\@ \cite{Wong2022TB-Net:Images} introduced TB-Net, a novel architecture specifically tailored to TB detection in CXRs, which included self-attention mechanisms and explored new approaches to CNN-based TB detection. Despite previous efforts, considering the WHO recommendation for TB screening and the increase in undiagnosed and untreated TB, there is an urgent need to create new solutions specifically tailored for deployment in low-resource settings, where efficiency and computational time could be a challenge.

In this context, we propose LightTBNet, a novel light-weighted, fast and efficient CNN specially customized to detect TB from CXRs. The main objectives of our design have been to outperform current state-of-the-art methods while reducing computational and memory requirements, through a careful design of the number of residual blocks and skip connections, and by proposing a specific architecture that reduces the number of operations and thus increase efficiency. This could allow our method to be deployed on handheld devices maximizing accuracy. Additionally, we present a thorough assessment of the performance and efficiency of our method considering public datasets and state-of-the-art alternatives.   

\section{Methods}
\label{sec:methods}

\subsection{Dataset and splits}
\label{ssec:dataset}

For the development of this study, we considered a total of 800 frontal CXR images from two public datasets \cite{Jaeger2014TwoDiseases}:

\begin{itemize}
  \item Montgomery County X-ray Set (MC): 138 frontal CXRs. 80 CXRs are non-TB and 58 are abnormal with manifestations of TB, including effussions and milary patterns.
  \item Shenzhen Hospital X-ray Set (SZ): 662 frontal CXRs. 326 CXRs are non-TB and 336 are abnormal showing diverse manifestations of TB, including some pediatric CXRs.
\end{itemize}

\noindent
The mean age of the whole set (MC \& SZ) is of 36.24 $\pm$ 15.65 y.o. (MC: 40.11 $\pm$ 18.72 y.o., SZ: 35.43 $\pm$ 14.80 y.o.). The dataset was randomly split following a training-validation-testing scheme. Twenty percent of the data was reserved for independent testing, with no patient overlap, and stratifying the data by cohort, TB class (positive/negative), sex and age, ensuring that the data were as balanced as possible. The remaining 80\% was further split using a 5-fold cross-validation approach for hyperparameter tuning.

\subsection{Image preprocessing and data augmentation}
\label{ssec:preprocess}

We applied Contrast Limited Adaptive Histogram Equalization (CLAHE) to all the data, as it has been shown to improve the detection of TB in CXRs \cite{Zuiderveld1994ContrastEqualization, Lakhani2017DeepNetworks, Siracusano2020Pipelineclahe}. The images were downscaled to 256$\times$256 when inputted to the networks, ensuring that the tensors are not excessively large while keeping the spatial resolution required to distinguish TB findings, if present.

The training data were augmented with a random horizontal flip, a random rotation from -15º to +15º, a random shift from -10\% to +10\% (both height and weight) and a random scaling from -10\% to +10\%. The input images were normalized to zero mean and unit variance in order to train the network more efficiently.

\subsection{Model architecture}
\label{ssec:model-architecture}

\begin{figure*}[htbp]
    \centering
    \includegraphics[width=0.85\textwidth]{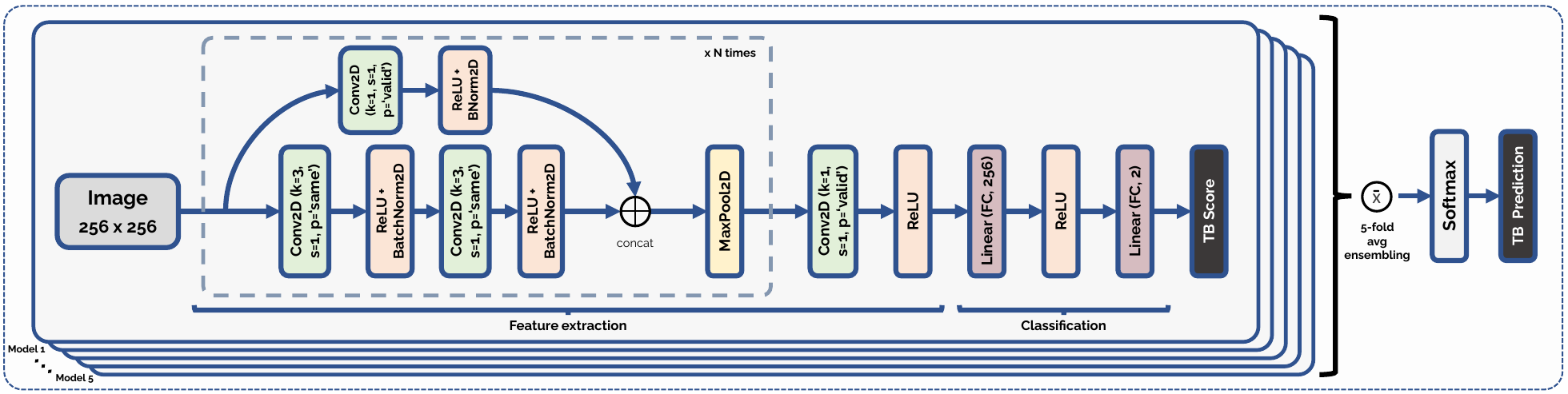}
    \caption{Model architecture (LightTBNet) and pipeline for TB prediction. $N$ corresponds to the hyperparameter that controls the number of residual blocks included in the model. Inference is done by ensembling five models obtained from five different cross-validation folds.}
    \label{fig:architecture}
\end{figure*}

The proposed architecture (see Figure \ref{fig:architecture}) is an efficient, lightweight adaptation of the ResNet architecture composed of $N$ residual convolutional blocks. These residual blocks consist of two 3$\times$3 convolutions followed by ReLU activations and batch normalizations (BatchNorm). Parallel to these two convolutions, there are skip connections with a 1$\times$1 convolution, which act as shortcuts when the tensors of both branches are concatenated. The residual block ends with a max-pooling layer (pooling size 2$\times$2, stride 2). These residual blocks are followed by a 1$\times$1 convolutional layer, which reduces the dimensionality of the feature maps extracted from the previous blocks. Then, an MLP-based classification module composed of two fully-connected layers is included. 

There is a trade-off between inference time and model size that depends on the number of residual blocks ($N$). The more residual blocks we include to the architecture, the lighter the model will be, since the extracted feature maps will be smaller. However, this leads to an increase in the number of multiply–accumulate operations (MACs) and inference time. Moreover, the use of skip connections in the residual blocks allows the model to automatically decide where to pass the information through, following an optimal pathway for the given task. The number of convolutional layers impacts on the detection of TB findings in CXR images, since the more convolutional layers the model has, the more general features it will extract from the image.

The design of our architecture differs from traditional ResNets in the introduction of pooling layers between convolutions and in the use of convolutional steps within the skip connections. In addition, our proposal includes a fewer number of convolutional layers and a smaller number of parameters. The use of fewer convolutional layers, ReLU activation functions and max-pooling layers (which reduce tensor dimensionality) within the architecture leads to a fewer number of MACs and parameters, thus making it more efficient.

\begin{table*}[htbp]
\centering
\resizebox{0.85\textwidth}{!}{%
\begin{tabular}{@{}llllllll@{}}
\toprule
\textbf{Model} & \textbf{Val ACC} & \textbf{Val F1} & \textbf{Val AUC} & \textbf{MACs (G)} & \textbf{Params (M)} & \textbf{Inference time (ms)} & \textbf{Size (MB)} \\ \midrule
\textbf{LightTBNet (N=3)} & \textbf{0.869 ± 0.024} & 0.874 ± 0.027 & 0.948 ± 0.016 & \textbf{0.822} & 4.298 & \textbf{2.25 ± 0.63} & 150.02 \\
\textbf{LightTBNet (N=4)} & 0.867 ± 0.057 & 0.873 ± 0.043 & \textbf{0.952 ± 0.02} & 1.138 & \textbf{1.467} & \textbf{2.55 ± 0.21} & \textbf{148.44} \\
\textbf{LightTBNet (N=5)} & 0.836 ± 0.04 & \textbf{0.878 ± 0.019} & 0.947 ± 0.013 & 1.456 & 1.932 & 3.22 ± 0.27 & 154.9 \\ \bottomrule
\end{tabular}%
}
\caption{Results for model comparison in validation set (cross-validation). The inference time is expressed in milliseconds (ms) and the values correspond to the time taken to predict one image by a single fold model. The model size is expressed in megabytes (MB) and corresponds to the size of a single fold model (including architecture and weights). The best option in each metric is highlighted in bold.}
\label{tab:results-cv}
\vspace{-0.4cm}
\end{table*}

Other high-performing and efficient architectures including DenseNet-121 (used by Rajpurkar et al.\@ \cite{Rajpurkar2020CheXaid:HIV}), EfficientNet (B0 and V2-s), MobileNet (MobileNetV3-small) and ResNet (ResNet-18, ResNet-34), were used in the experiments to compare with the performance and efficiency of our proposal.

\section{Experiments and results}
\label{sec:experiments_results}

\subsection{Implementation details}
\label{ssec:implementation_details}

For all the experiments, we used Adam optimizer with default parameters ($\beta_{1} = 0.9, \beta_{2} = 0.999$), batch size of 16 images, learning rate $1\times~e^{-4}$, and focal loss \cite{Lin2020FocalDetection}, a variation of the cross entropy loss which introduces a modulating term $(1-p_{t})^{\gamma}$ that downweights the loss assigned to well-classified examples (easy examples) and upweights the loss assigned to hard-to-classify examples, focusing learning on negative hard examples. The focal loss is formulated as follows:

\begin{equation}
    FL~(p_{t}) = -(1-p_{t})^{\gamma}~log(p_{t})
\end{equation}

where $p \in [0,1]$ is the model's estimated probability for target label $t$ and $\gamma \geq 0$ is a tunable focusing parameter.

We also performed 5-fold cross-validation. On each fold, we evaluated performance on the validation subset and chose the best model checkpoint based on the highest validation AUC.

During the testing of the models, we used the best algorithm trained on each fold and averaged their predictions (TB scores) across an ensemble of the five models (see Figure \ref{fig:architecture}). Previous experiments showed better performance when ensembling the five models extracted from each fold, ensuring a more robust final score.

To implement the networks, we used the PyTorch framework (v.\@ 1.12). For the experiments, we used a workstation with NVIDIA A30 24GB GPU, 256GB RAM and 2 x Intel Xeon Silver 4216 @ 2.1 GHz CPUs.

\subsection{Hyperparameter optimization}
\label{ssec:hyperparameter}
We compared the TB classification performance, efficiency and size among three different versions of the proposed architecture, in which we vary the the number of residual convolutional blocks ($N$). We considered different architectures ($N$=3,4,5) and selected the best performing configuration on the cross-validation subsets by analyzing their AUC scores. We compared these three versions since, if we took a smaller number of $N$, the network was not deep enough to correctly find TB-compatible findings, and if we took a higher value, the feature maps extracted after the convolutional steps were extremely small considering the input image size, and there was a significant loss of spatial information in the process.

We also calculated the multiply–accumulate operations (MACs)\footnote{\href{https://github.com/Lyken17/pytorch-OpCounter}{Code used: https://github.com/Lyken17/pytorch-OpCounter}}, the number of parameters and inference time for each of the configurations in order to compare their efficiency and computational requirements. These calculations were performed by running 300 repetitions in a row and then averaging the timings, thus preventing the GPU from going into power-saving mode when measuring time.

The results obtained for each of the configurations in the validation set are included in Table \ref{tab:results-cv}. All three versions demonstrated high efficiency and low inference times. LightTBNet with 4 residual convolutional blocks ($N=4$) shows better performance (val AUC) with low computational requirements when comparing with the other configurations ($N$ = 3, 5) tested. As a result, this configuration was selected for testing.

\subsection{Results on test set}
\label{ssec:tb_detection}

\begin{figure}[htbp]
\vspace{-0.2cm}
    \centering
    \includegraphics[width=0.45\textwidth]{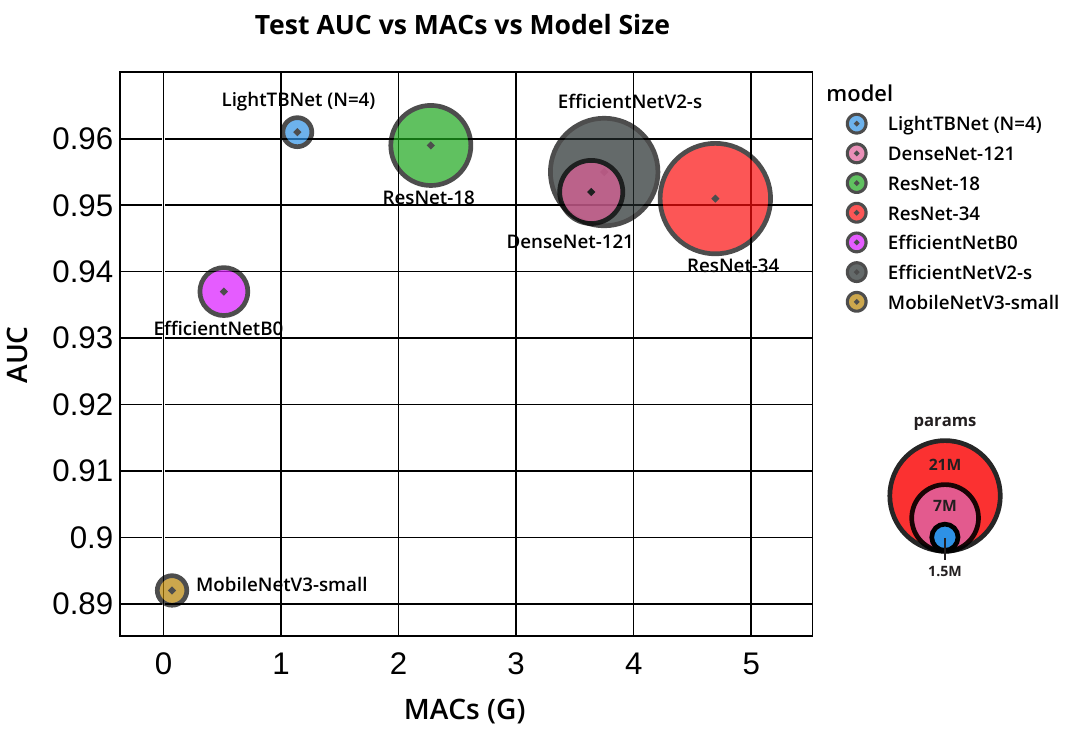}
    \captionsetup{skip=4pt}
    \caption{Performance (AUC score) on the test set vs.\@ efficiency (MACs \& number of parameters). Comparison between our proposal and other high-performing and efficient architectures.}
    \label{fig:mac-auc-test-graph}
\vspace{-0.5cm}
\end{figure}

In this section, we test our best configuration ($N$=4) along with other high-performing and efficient architectures, detailed in section \ref{ssec:model-architecture}, on an independent test subset of 160 images. The Table \ref{tab:results-comparison-test} shows that our proposal outperforms the results of other efficient approaches both in terms of performance (highest AUC score) and efficiency (lowest number of parameters and inference time). The inference times of all the architectures were obtained following the same procedure as detailed in section \ref{ssec:hyperparameter}. Moreover, a graphical comparison taking into account performance (AUC score) and efficiency (MACs \& number of parameters) of each of the architectures, is shown in Figure \ref{fig:mac-auc-test-graph}.

\begin{table*}[htbp]
\centering
\resizebox{0.76\textwidth}{!}{%
\begin{tabular}{@{}llllllll@{}}
\toprule
\textbf{Model} & \textbf{Test ACC} & \textbf{Test F1} & \textbf{Test AUC} & \textbf{MACs (G)} & \textbf{Params (M)} & \textbf{Inference time (ms)} & \textbf{Size (MB)} \\ \midrule
\textbf{DenseNet-121} & \textbf{0.906} & \textbf{0.908} & 0.952 & 3.639 & 6.95 & 26.09 ± 0.9 & 411.01 \\
\textbf{EfficientNetB0} & 0.844 & 0.839 & 0.937 & 0.513 & 4.01 & 13.23 ± 0.33 & 213.51 \\
\textbf{EfficientNetV2-s} & 0.9 & 0.893 & 0.955 & 3.749 & 20.18 & 27.7 ± 0.69 & 444.82 \\
\textbf{MobileNetv3-small} & 0.812 & 0.803 & 0.892 & \textbf{0.072} & 1.52 & 6.59 ± 0.43 & 46.93 \\
\textbf{ResNet-18} & 0.875 & 0.868 & 0.959 & 2.274 & 11.176 & 3.79 ± 0.76 & \textbf{118.64} \\
\textbf{ResNet-34} & 0.9 & 0.896 & 0.951 & 4.695 & 21.288 & 7.56 ± 0.32 & 207.21 \\
\textbf{LightTBNet (N=4)} & \textbf{0.906} & 0.907 & \textbf{0.961} & 1.138 & \textbf{1.467} & \textbf{2.55 ± 0.21} & 148.44 \\ \bottomrule
\end{tabular}%
}
\caption{Results for model comparison in test set. The inference time is expressed in milliseconds (ms) and the values correspond to the time taken to predict one image by a single fold model. The model size is expressed in megabytes (MB) and corresponds to the size of a single fold model (including architecture and weights). The best option in each metric is highlighted in bold.}
\label{tab:results-comparison-test}
\end{table*}

\begin{table*}[htbp]
\centering
\resizebox{0.75\textwidth}{!}{%
\begin{tabular}{@{}lllllll@{}}
\toprule
\textbf{Model /   Implementation} & \textbf{ACC (MC + SZ)} & \textbf{AUC (MC + SZ)} & \textbf{ACC (MC)} & \textbf{AUC (MC)} & \textbf{ACC (SZ)} & \textbf{AUC (SZ)} \\ \midrule
\textbf{Hwang et al. \cite{Hwang2016ANetworks}} & - & - & 0.67 & 0.88 & 0.84 & 0.93 \\
\textbf{Islam et al. \cite{Islam2017AbnormalityNetworks}} & - & - & - & - & 0.88 & 0.91 \\
\textbf{Pasa et al. \cite{Pasa2019EfficientVisualization}} & 0.862 & 0.925 & 0.79 & 0.811 & 0.844 & 0.9 \\
\textbf{LightTBNet (N=4)} & \textbf{0.906} & \textbf{0.961} & \textbf{0.889} & \textbf{0.943} & \textbf{0.91} & \textbf{0.963} \\ \bottomrule
\end{tabular}%
}
\caption{Accuracy (ACC) and AUC results of our model on the Montgomery County (MC) and Shenzhen (SZ) test subsets in comparison to the results reported in other publications. To provide a better comparison with our results, figures from other publications where pre-training is performed on ImageNet and/or ensembles are used have been discarded. Only deep learning-based approaches with results on MC \& SZ were considered. The best option in each metric is highlighted in bold.}
\label{tab:comparison-papers}
\end{table*}

\begin{figure}[htbp]
    \centering
    \includegraphics[width=0.45\textwidth]{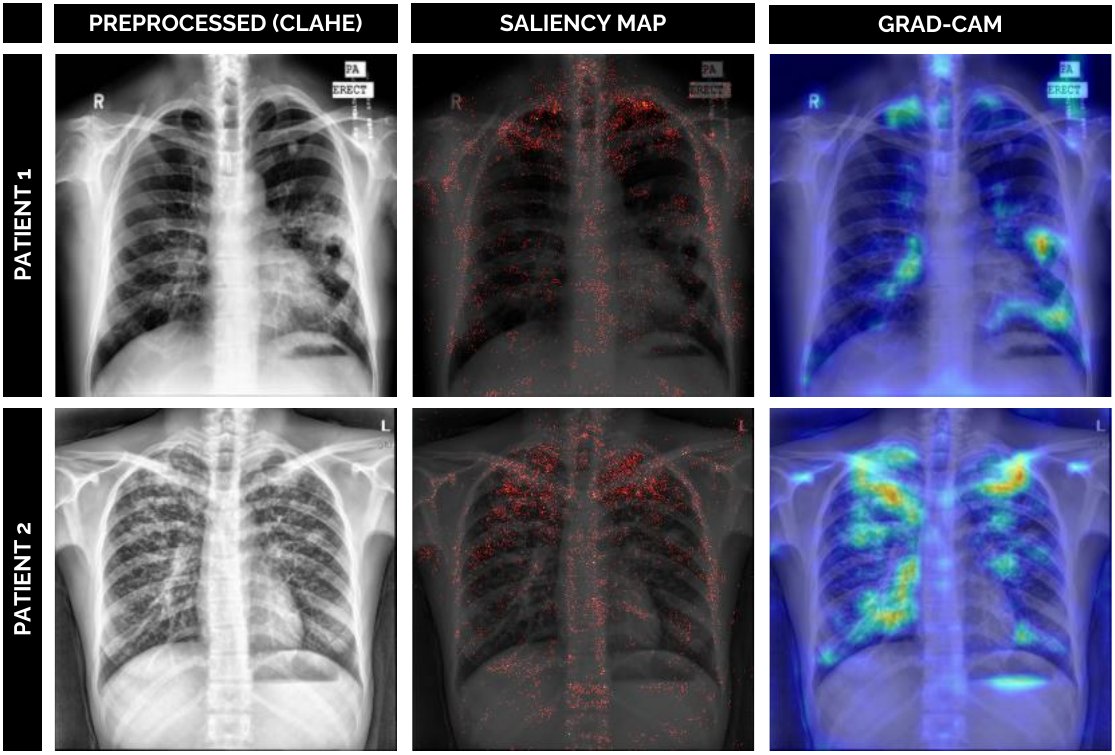}
    \captionsetup{skip=4pt}
    \caption{Preprocessed image (CLAHE), saliency map and grad-CAM of two TB positive cases truly predicted as positive. Patient 1 belongs to MC and patient 2 to SZ. The output TB scores provided by the model were 0.8215 and 1.0 for patient 1 and 2, respectively.}
    \vspace{-0.4cm}
    \label{fig:saliency_gradcam}
\end{figure}

Next, we compared as fairly as possible our proposal with other approaches in the literature that report results on the same TB datasets. To do so, we considered different deep-learning proposals and selected the best results from each contribution which are comparable to our results. As shown in Table \ref{tab:comparison-papers}, our proposal outperforms the results of other contributions, both when combining and splitting separately the two TB datasets. However, our work could not be directly compared to other efficient TB methodologies, such as those proposed by Wong et al.\@ (TB-Net) \cite{Wong2022TB-Net:Images}  and Lakhani et al.\@ \cite{Lakhani2017DeepNetworks}, as their results were not reported independently for the two public datasests used in this work. When testing on the independent test set, the proposed model provides an ACC, F1 and AUC of 0.906, 0.907 and 0.961, respectively. With a default prediction threshold of $0.5$, we achieve a sensitivity (SN) and specificity (SP) of 0.924 and 0.889, respectively, thus fulfilling the requirements of the WHO's Target Product Profile (TPP) of triage tests ($\geq$90\% SN and $\geq$70\% SP). Finally, Figure \ref{fig:saliency_gradcam} displays the saliency maps and grad-CAMs of two true positive (TP) cases, i.e., truly classified as positive for TB. Saliency maps can be used to assess an image's overall spatial support for a given class. On the other hand, grad-CAMs are coarse location maps that highlight the key areas of the image that the model focuses on for prediction. As shown in Figure \ref{fig:saliency_gradcam}, the proposed model correctly identifies a TB-compatible cavity in the left lung of patient 1 and shows a generalized activation in the image of patient 2 due to the miliary pattern present in the CXR.

\section{Conclusion}
\label{sec:conclusion}

In this study, we have proposed a lightweight, fast and efficient DL-based residual network for detecting TB from CXRs, which has shown outstanding performance and low computational requirements compared to other state-of-the-art architectures and implementations. Given these low computational requirements and good performance, this solution demonstrates its potential use in computer-aided diagnosis systems and could be deployed for use in smartphones or other handheld devices, serving to gain accessibility to diagnosis and to reduce the high clinical burden when screening for TB, which remains a challenging task, especially in low-resource settings with high TB prevalence.

\newpage
\section{Compliance with ethical standards}
\label{sec:ethics}

\vspace{0.2cm}
\noindent
This research study was conducted using retrospective data from open access sources (details in text). Ethical approval was not required as confirmed by the license of the public databases used.

\section{Acknowledgements}
\label{sec:acks}
%
This work was supported by H2020-MSCA-RISE-2018 INNOVA4TB (EU) project (ID 823854) and ADVANCE-TB Cost Action (CA21164 (EU)). DCM's PhD fellowship was supported by Universidad Politécnica de Madrid.

\bibliographystyle{IEEEbib-abbrev-etal}
\bibliography{references}

\end{document}